\documentstyle[12pt,thmsa,sw20aip]{article}

\input tcilatex
\begin{document}

\title{}

Topological Gauge Theory Of General Weitzenb\"{o}ck Manifolds Of
Dislocations In Crystals

Y. C. Huang$^{1,3}$ B. L. Lin$^2$ S. Li$^3$ M. X. Shao$^{3,4}$

$^1$Department of Applied Physics, Beijing Polytechnic University,
Beijing,100022, P. R. China

$^2$Department of Transportation Management Engenerring, Northean Jiaotong
University, Beijing, 100044, P. R.China

$^3$Institute of Theoretical Physics, Chinese Academy of Science, Beijing,
100080, P. R. China

$^4$Department of Physics, Beijing Normal University, Beijing, 100022, P. R.
China

( $^1$Email: ychuang@bjpu.edu.cn )

Abstract

General Weitzenb\"{o}ck material manifolds of dislocations in crystals are
proposed, the reference, idealized and deformation states of the bodies in
general case are unifiedly described by the general manifolds, the
topological gauge field theory of dislocations is given in general case,
true distributions and evolution of dislocations in crystals are given by
the formulas describing dislocations in terms of the general manifolds,
furthermore, their properties are discussed.

1.Introduction

Practical crystals always contain a lot of dislocations and are filled with
them, the defects strongly affect the properties of the crystals. Some
pioneers even in the fifties had researched the relations between the defect
crystals and differential geometry. K. Kondo used torsion of material
manifolds to study dislocations$^1$, and Nye$^2$, Bilby$^3$, Kr\"{o}ner$^4$
etc studied the relations between the defects and torsion.

On the other hand, these defects are found having the features of elementary
particles, and Kr\"{o}ner studied continuum theory of defects$^5$. Due to
the great success of Yang-Mills gauge field theory describing the
interactions of elementary particles$^6$, according to the research of gauge
theory of gravitation, it can be discovered that geometry of Riemannian
manifold essentially belongs to a kind of non-Abelian gauge theory$^7$.
Gauge theory and topological properties of dislocations were researched, and
many important conclusions in the fields were obtained by Toulous and Kleman$%
^8$, Kleinert$^9$, Trebin$^{10}$ and so on. Duan and his associates showed a
unified approach to the study of defect mechanics and their topological and
geometric properties by applying vielbein field theory and gauge field
theory to dislocation and disclination continuum$^{11,12}$.

Sect.2 of the paper constructs the general Weitzenb\"{o}ck manifolds of
dislocations in crystals, gives a unified description of these reference,
idealized and deformation states; Sect.3 unifiedly studies topological gauge
theory of dislocations, gives the formulas describing dislocations, their
properties are discussed; Sect.4 gives the discussions of overcoming the
default; the last Sect. is summary.

2.General Weitzenb\"{o}ck Material Manifolds of Dislocations

The Weitzenb\"{o}ck manifold$^7$is the one whose $\nabla _ig_{jk}=0$, $%
T_{jk}^i\neq 0$, $R_{jkl}^i=0$, where $g_{ij}$, $T_{jk}^i$ and $R_{jkl}^i$
are, respectively, metric, torsion and curvature tensors of the manifold,
when $T_{jk}^i=0$, the manifold is reduced into Euclidean manifold. Since
properties of dislocations correspond to properties of torsion tensor of a
manifold$^{1-4}$, when considering dislocations, the corresponding manifold
is thus taken as the Weitzenbock manifold. And the torsion tensor and metric
tensor $g_{ij}$ of a manifold are all defined locally on the manifold$^{13}$%
, then we may naturally expect that the torsion tensor possesses different
values, even zero, on different parts of the manifold, we call the manifold
as the general Weitzenbock manifold, the corresponding different parts are
just the corresponding submanifolds in the material manifold. In fact, the
situations above just correspond to disappearance and appearance of
dislocations for zero and non-zero of torsion tensor of the corresponding
submanifold respectively. Therefore, we can use the above discussions to
represent the complex situations of motion, disappearance and appearance of
the dislocations.

Now we construct the general Weitzenb\"{o}ck material manifolds, relative to
time paeamotor t, of dislocations in crystals.

It is well known that Weitzenbock manifold is an ensemble of lots of small
coordinate pieces homeomorphic to some local Euclidean spaces, lying,
however, to a certain degree amorphously, and the characters of the ensemble
of lots of the small pieces can be described by means of affine connections
of the small pieces. And the geometric properties of dislocations
homeomorphically correspond to the geometric properties of torsions of the
corresponding material manifold.

Owing to the periodic regular distributions of the lattice particles of
perfect crystal, the position coordinate $x^a$ of lattice particles can be
globally determined in Euclidean space in terms of their periodic rules,
then we regard their corresponding material manifold as Euclidean manifold.
Obviously, when dislocations appear, which means that we can not globally do
the same in terms of their periodic rules, however, we can build up a lot of
local coordinate systems in which the position coordinate $x^a$ of any
particles in the crystal can be locally defined.

In general, for a crystal with dislocations in 3-dimensional Euclidean
space, we assume that the body is made of N lattice particles, for any
lattice particle $P_{ia}$, we can always construct local coordinate system $%
L_{ia}$ in local Euclidean space and presume that there are k lattice
particles $P_{i1},P_{i2},\cdot \cdot \cdot \cdot \cdot \cdot ,P_{ik}$ which
are close adjacent to $P_{io}$, and the corresponding nonholonomic
normalized frame fields are ${\bf e}^a(a=1,2,3)$, then the distance vector
of any two close adjacent particles is 
\begin{equation}
dr=e^adx^a
\end{equation}
the metric tensor is 
\begin{equation}
\eta ^{ab}=e^a\cdot e^b=\delta ^{ab}
\end{equation}
and the square of the distance, i.e., the square of the line element is 
\begin{equation}
ds^2=dr\cdot dr=\eta ^{ab}dx^adx^b=dx^adx^a
\end{equation}
where $dx^a$ is the corresponding nonholonomic coordinate difference.

Furthermore, for any particle $P_{il}(l\neq 0)$ discussed above, we can
again construct a local coordinate system, and regard any such $P_{il}(l\neq
0)$ as a new $P_{io}^{^{\prime }},$ do further the same as the above
discussions about the $P_{il}$, $\ldots \ldots ,$ up to all particles in the
body.

For a open subset $V_\alpha \subset R^3$ of any local coordinate system
discussed above, assume that there is an inverse homeomorphic map $\varphi
_\alpha ^{-1}:$ $V_\alpha \rightarrow M_\alpha $ ( where $M_\alpha $ is a
family of open sets ) and $\bigcup\limits_\alpha V_\alpha =V_m$ ( $V_m$ is
the volume of the crystal in Euclidean space ) such that $%
\bigcup\limits_\alpha M_\alpha =M$ is topological space, which means that M
is provided with family of pairs $\{(M_\alpha ,\varphi _\alpha )\}$. Further
assume that given $M_\alpha $ and $M_\beta (\alpha \neq \beta )$ ( $M_\alpha
\bigcap $ $M_\beta \neq \phi $ ) the map $\varphi _\beta \circ \varphi
_\alpha ^{-1}$from the subset $\varphi _\alpha (M_\alpha \bigcap M_\beta )$
of $R^3$ to the subset $\varphi _\beta (M_\alpha \bigcap M_\beta )$ of $R^3$
is infinitely differentiable. Consequently, the aggregation of the all
coordinate pieces forms a general three dimensional differentiable material
manifold$^{13}$. Because the general manifold inherits the topological and
geometric properties of the dislocations in the body, and these properties
correspond to properties of torsion of the manifold and can just be
expressed by the torsion tensor of the manifold, therefore, there always
exists the inverse map $\varphi _\alpha ^{-1}$ such that M is the manifold
that the properties of the dislocations of the crystal may be represented by
means of the torsion tensors of the manifold. Because the torsion of the
general manifold may have different values on different parts of the
manifold, the different parts may correspond to Euclidean and
Weitzenb\"{o}ck material submanifolds of the general material manifold.
Evidently, the local Euclidean and Weitzenb\"{o}ck material submanifolds can
be locally included in the general Weitzenb\"{o}ck material manifold, which
just corresponds to the real distribution situations of dislocations of the
body.

The relationship$^{15,16}$ between the normalized nonholonomic and holonomic
frames is 
\begin{equation}
e^a=e_i^ae^i,a,i=1,2,3
\end{equation}
where $e^i=dx^i$ is holonomic coframe of the manifold and $e_i^a$ is the
vielbein, then we have metric tensor of the manifold $M$%
\begin{equation}
g_{ij}=e_i^ae_j^a
\end{equation}

Using the vielbein theory$^{7,14-16}$ we can have 
\begin{equation}
dx^a=e_i^adx^i
\end{equation}

Substituting (6) into (3) and using (5), we have 
\begin{equation}
ds^2=dx^adx^a=e_i^adx^ie_j^adx^j=g_{ij}dx^idx^j
\end{equation}
which means that the square of the line element of a local Euclidean system
in the crystal is equal to that of the line element of the general material
manifold. Using the above discussions, furthermore, we can study their
stress and so on. Therefore, the above discussions is essential for our
further study.

If we take time $t_0$ to label the reference state of time $t_0$, when the
body is experienced a deformation during time $\triangle t=t-t_0$ the
corresponding deformed state is labelled by time $t=t_0+\triangle t$, in
other words, at different time $t$, the lattice particles possess different
distributions by which the creating and moving of the dislocations with time 
$t$ are represented, the general material manifold with time parameter t is
thus obtained.

In elastic and plastic mechanics, the reference, idealized and deformation
states are chosen. We now give a unified description of the states by means
of the general manifold. The reference state is usually chosen as perfect or
idealized states of an ordered material. By the discussions above of
translating the discretum of a perfect crystal into the continuum, we obtain
a Euclidean material manifold corresponding to the global vanishing torsion
of the manifold. Usually the idealized state of the body is defined as an
aggregation of lots of small pieces of the idealized material. In fact, the
small pieces of the idealized material can always be viewed as having some
ordered distributions of lattice particles, and our discussions about the
piece distributions of the particles are general, i. e., our discussions are
general, then the idealized state of the body is included in the states of
the general Weitzenb\"{o}ck material manifold M. For elastic and plastic
deformation states, they can be represented in microscope as the changes of
the displacement and interaction of any adjacent lattice particles.
Analogous to the above discussions, evidently the elastic and plastic
deformation states are included in the general state of the general manifold
M , especially plastic deformation has cooperative motions of the many
defects and lattice particles. Now we generally consider reference state (or
called standard state ) and deformation state of M , the line elements of
their manifolds are generally defined, respectively, as 
\begin{equation}
ds_r^2=ds_{t_0}^2=g_{ij}(x(t_0),t_0)dx^i(t_0)dx^j(t_0)
\end{equation}
and 
\begin{equation}
ds_d^2=ds_t^2=g_{ij}(x(t),t)dx^i(t)dx^j(t)
\end{equation}

Using (1) and (5), Eq.(9) can be generally rewritten as 
\begin{equation}
ds_t^2=\eta ^{ab}e_i^a(x(t),t)e_j^b(x(t),t)dx^i(t)dx^j(t)
\end{equation}

Similar to Ref.[11], the strain tensor is defined as 
\begin{equation}
E_{ij}=E_{ij}(x(t),x(t_0),t,t_0)=\frac 12[g_{ij}(x(t),t)-g_{ij}(x(t_0),t_0)]
\end{equation}

The general reference and deformation states and the strain tensor at any
time t are useful for practical calculation, and it is convenient for
acquiring different properties of a crystal, in fact, one usually compares
reference states with deformation states under different conditions and
studies the evolution laws of the deformation states with time t.

It is important and useful to describe these different states in general
case, we overcomes the difficulty of self- consistence that generally
describes these different states in terms of differential geometry, give
true distributions and evolution of dislocations with variances of time in
the crystals by means of the differential manifolds, and may give useful
tool of describing the defects.

3.Topological Gauge Field Theory Of Dislocations

We now discuss the topological gauge field theory of the defects. In
physics, any physical law doesn't depend on the choices of coordinates. For
example, the square of the line element and Lagrangian of the system, there
exist two kinds of indices i and a ( i, a, = 1, 2, 3 ) which is holonomic
and nonhonolomic indices respectively. For holonomic indices i, the
coordinate transformations at a certain time t, i.e ., at a certain state,
are 
\begin{equation}
x^{^{\prime }i}(t)=x^{^{\prime }i}(\{x^j(t)\})
\end{equation}
and 
\begin{equation}
x^i(t)=x^i(\{x^{^{\prime }j}(t)\})
\end{equation}

The transformation of the nonholonomic indices is the local SO(3) gauge
transformation$^{11}$%
\begin{equation}
dx^{^{\prime }a}(t)=S^{ab}(t)dx^b(t)
\end{equation}
where 
\begin{equation}
\eta ^{^{\prime }ab}=\eta ^{cd}S^{ac}S^{bd}
\end{equation}

Using Eqs.(14), (15) and the discussions above, it is easy to prove that
Eqs.(8-10) are all invariant under the two kinds of coordinate
transformations.

In order to achieve the invariance of physical laws under the two kinds of
coordinate transformations, the usual partial derivative should be
substituted into two kinds of covariant derivatives as follows 
\begin{equation}
\nabla _ie_j^a=\partial _ie_j^a-\Gamma _{ij}^ke_k^a
\end{equation}
and 
\begin{equation}
D_ie_j^a=\partial _ie_j^a-\omega _i^{ab}e_j^b
\end{equation}
where $\Gamma _{ij}^k$ and $\omega _i^{ab}$are affine connection and $SO(3)$
gauge potential (spin connection). Eqs.(16) and (17) satisfy the general and
gauge covariant principles respectively, but the total Lagrangian of the
system is invariant under the both transformations$^{16}.$

Using the vielbein theory$^{11,12,16}$ we have the torsion tensor of the
general manifold as follows: 
\begin{equation}
T_{ij}^k=\Gamma _{ij}^k-\Gamma
_{ji}^k=e^{ka}(D_ie_j^a-D_je_i^a)=e^{ka}T_{ij}^a
\end{equation}
where $T_{ij}^a$ is the torsion tensor with nonholonomic superscript a 
\begin{equation}
T_{ij}^a=D_ie_j^a-D_je_i^a
\end{equation}

Analogous to Ref.[12,17], we may define the general tensor densities of
dislocations of the body at any time t in the following 
\begin{equation}
\alpha ^{ia}(x(t),t)=\frac{\varepsilon ^{ilm}}{2\sqrt{g(x(t),t)}}%
T_{lm}^a(x(t),t)
\end{equation}
where $g=det(g_{ij}),$ $\varepsilon ^{123}=-\varepsilon ^{132}=1$, the
coefficient $\frac 12$ is due to the tensor sums of subscripts l and m .
Then Burgers vector of dislocation can be defined as follows 
\begin{equation}
b^a=\int_\Sigma \alpha ^{ia}(x(t),t)\sqrt{g(x(t),t)}d\sigma _i
\end{equation}
where $\Sigma $ is the surface including dislocations. The formulas (20) and
(21) and x(t) are all the functions of the time t that labels the motions of
the defects and deformations of a crystal, which includes the creations,
motions and disappearances of the dislocations with evolution of time t in
the general manifold, therefore, these variances can be represented by
Burgers vector, or the tensor density of dislocations, these variances
correspond to the variances that Eq.(20) and Eq.(21) are equal to different
values at different time and position. Then it is given that true
distributions and evolution of dislocations in the crystals by the formulas
describing dislocations at any time t in the general manifold, and our
theory is non-linear. Using gauge potential decompositions Duan and Zhang$%
^{12}$ gave the moving, geometric and topological descriptions of
dislocations in terms of similar to Eqs. (20) and (21). Since the general
Weitzenbock manifold is very general, then the corresponding further
discussions of (20) and (21) are similar to those in Ref.[12], therefore,
one can arrive in their all results of topology, geometry and kinematics of
the defects, thus we shall not repeat here.

The discussions about elastic and plastic dynamics and the generalization to
a four dimensional general pseudo-Weitzenbock material manifold that
concerns dissipative parts belonging to the fourth component of the vielbein
fields will be written in a separated paper.

4.Summary

This paper proposes general Weitzenb\"{o}ck material manifolds, relative to
time parameter t, of dislocations in crystals, in the manifolds the local
Euclidean and Weitzenb\"{o}ck manifolds can be viewed as the submanifolds of
the general Weitzenb\"{o}ck material manifold, furthermore, by analyzing the
microscope structures and change, with time t, of the general manifolds, we
obtain the unified description of reference, idealized and deformation
states in terms of the general manifolds, and the general manifold with
evolution of time parameter t just represents the motions and the
distributions of dislocations, and the topological gauge theory are acquired
by leaving the physical laws invariant under the coordinate transformations.
We overcome the difficulty of self-consistence that generally describes
these different states in terms of differential geometry, give true
distributions and evolution of dislocations in the crystals by the formulas
describing dislocations to any time in general manifold, and their
properties are discussed.

Acknowledgment

One, Huang, of the authors is grateful to Prof. Y. S. Duan and Prof. S. L.
Zhang for help.

\end{document}